\pdfoutput=1

\documentclass[twocolumn]{aastex631}

\usepackage{xspace}
\usepackage{ulem}

\newcommand{\x}[2]{\textcolor{purple}{\sout{#1}\bf #2}}
\newcommand{\Ms}{{\ensuremath{\mathrm{M}_{\odot}}}\xspace}

\usepackage{amsmath}
\usepackage{epsfig}
\usepackage{amssymb}
\usepackage{graphicx}
\usepackage{color}
\usepackage{natbib}
\usepackage{ulem}
\usepackage[usestackEOL]{stackengine}

\shorttitle{Multiple Supermassive Stars}
\shortauthors{Woods et al.}
\graphicspath{{./}{figures/}}

\begin{document}

\title{On the Formation and Interaction of Multiple Supermassive Stars in Cosmological Flows}

\correspondingauthor{Tyrone E. Woods}
\email{tyrone.woods@nrc-cnrc.gc.ca}

\author[0000-0003-1428-5775]{Tyrone E. Woods}
\affiliation{National Research Council of Canada, Herzberg Astronomy \& Astrophysics Research Centre, 5071 West Saanich Road, Victoria, BC V9E 2E7, Canada}
\affiliation{Monash Centre for Astrophysics, School of Physics and Astronomy, Monash University, VIC 3800, Australia}

\author[0000-0002-5293-699X]{Samuel Patrick}
\affiliation{Institute of Cosmology and Gravitation, University of Portsmouth, Portsmouth PO1 3FX, UK}

\author[0000-0002-1463-267X]{Daniel J. Whalen}
\affiliation{Institute of Cosmology and Gravitation, University of Portsmouth, Portsmouth PO1 3FX, UK}

\author[0000-0002-3684-1325]{Alexander Heger}
\affiliation{Monash Centre for Astrophysics, School of Physics and Astronomy, Monash University, VIC 3800, Australia}
\affiliation{ARC Centre of Excellence for Gravitational Wave Discovery (OzGrav), Melbourne, Australia}
\affiliation{ARC Centre of Excellence for Astrophysics in Three Dimensions (ASTRO-3D), Australia}
\affiliation{Joint Institute for Nuclear Astrophysics, 1 Cyclotron Laboratory, National Superconducting Cyclotron Laboratory, Michigan State University, East Lansing, MI 48824-1321, USA}

\begin{abstract}

Supermassive primordial stars with masses exceeding $\mathord\sim10^5\,\Ms$ that form in atomically cooled halos are the leading candidates for the origin of high-redshift quasars with $z>6$. Recent numerical simulations, however, find that multiple accretion disks can form within a halo, each of which can host a supermassive star. Tidal interactions between the disks can gravitationally torque gas onto their respective stars and alter their evolution.  Later, when two satellite disks collide, the two stars can come into close proximity.  This may induce additional mass exchange between them. We investigate the co-evolution of supermassive stars in atomically-cooled halos driven by gravitational interactions between their disks.  We find a remarkable diversity of evolutionary outcomes.  The results depend on these interactions and how the formation and collapse times of the stars in the two disks are correlated.  They range from co-evolution as main sequence stars to main sequence -- black hole pairs and black hole -- black hole mergers.  We examine the evolution of these secondary supermassive stars in detail and discuss the prospects for binary interactions on much smaller scales after the disks merge within their host halos.

\end{abstract}


\section{Introduction} \label{sec:intro}

More than 200 quasars powered by supermassive black holes have now been discovered at $z>6$ \citep{quasardataset}, including nine at $z>7$ \citep{mort11, wu15, ban18,Wang21}.  A natural explanation for the seeds of these quasars are direct collapse black holes (DCBHs), which arise due to the formation of supermassive stars (SMSs) in atomically cooled halos at $z\gtrsim20$ \citep{bl03,ln06,wta08,rh09b,latif13a}.  In this picture, a primordial halo grows to $\mathord\sim10^7\,\Ms$ without forming stars because, for example, it is immersed in strong Lyman-Werner (LW) UV fluxes \citep{dijkstra08,agarw12,latif14} or highly supersonic streaming motions of gas relative to dark matter \citep{Tanaka2014,hir17,anna17}, or it is dynamically heated by violent mergers \citep{yahs03,Fernandez2014,wise19,Regan20}.  At this mass, virial temperatures reach $\mathord\sim10^4\,\mathrm{K}$ that activate line cooling in hydrogen that triggers rapid baryon collapse at initial rates of $0.1-1\,\Ms\,\mathrm{yr}^{-1}$.  Stellar evolution calculations indicate that such flows, if they persist, build up $10^4-10^5\,\Ms$ stars that later collapse to DCBHs (\citealt{hir13,um16,tyr17,hle18b} -- see \citealt{titans} and \citealt{maio19} for recent reviews).

Cosmological simulations of the collapse of atomically-cooled halos are either run at extremely high resolutions that can follow flows down to protostellar radii but only for short times because of time step restrictions \citep[e.g.,][]{bec15,bec18,ard18,luo18}, or for the longer times required to evolve the flows over many dynamical times but at the cost of resolving fragmentation deep in the accretion disk of the star \citep[e.g.,][]{chon18,rd18b,suaz19,latif21a}.  The first simulations to evolve atomically cooled flows for the entire life of a SMS were \citet{latif20a}, which showed that some could form in binaries or small clusters.  Most recently, \citet{pat20a} followed the collapse of a variety of atomically-cooled halos for $2-4\,\mathrm{Myr}$ and found that multiple accretion disks could form in a halo, each of which could harbor a SMS. These disks experienced multiple close encounters that gravitationally torqued gas into their centers, triggering brief bouts of massive accretion onto their respective stars.  These stars thus interacted indirectly with one another via encounters between their accretion disks.  Although the simulations suggested a variety of final outcomes for these objects, they did not evolve the stars themselves.

Recently, \citet{tyr21a} modeled the growth of the stars of the primary disks in \citet{pat20a} and found that they collapse at final masses of $1.1\times10^5\,\Ms-1.9\times10^5\,\Ms$.  Here, we investigate the co-evolution of the stars in the most massive and stable of the companion disks in \citet{pat20a} with the {\sc Kepler} Lagrangian stellar evolution and hydrodynamics code.  Our {\sc Kepler} models use the time-dependent accretion rates that were tallied at the centers of the disks in the cosmological simulations.  We follow the evolution of the SMSs in the companion disks until the disks were subsumed into the primary disks or torn apart by tidal interactions with them.  In Section 2 we describe our cosmological and stellar evolution models.  The evolution of the SMSs is examined in Section 3.  From the interaction histories of the disks and the formation and collapse times of the stars within them we determine if they form SMS -- SMS, SMS -- DCBH or binary DCBH systems and discuss possible consequences for their detection in Section 4.  We discuss prospects for binary interactions in these systems on much smaller scales at later times in Section 5.

\section{Numerical Methods} \label{sec:methods}

\subsection{\sc{enzo}}\label{sec:enzo}

Accretion rates for the companion stars in our study were taken from cosmological simulations with the {\sc enzo} adaptive mesh refinement code by \citet{pat20a}. These simulations resolved accretion at the centers of the disks on 0.01 pc scales with six-species primordial gas chemistry (H, He, e$^{-}$, H$^{+}$, He$^{+}$, and He$^{2+}$) to approximate very high LW backgrounds that ensured isothermal atomic cooling in the halos. The simulations included collisional ionization and excitation cooling by H and He, recombination cooling by H and He, bremsstrahlung cooling, and inverse Compton cooling by the cosmic microwave background \citep{enzo}.  The accretion rates were tabulated by computing mass fluxes through a 0.134 pc sphere at the center of each disk at 10 kyr intervals.

\begin{figure*}
    \includegraphics[width=\textwidth]{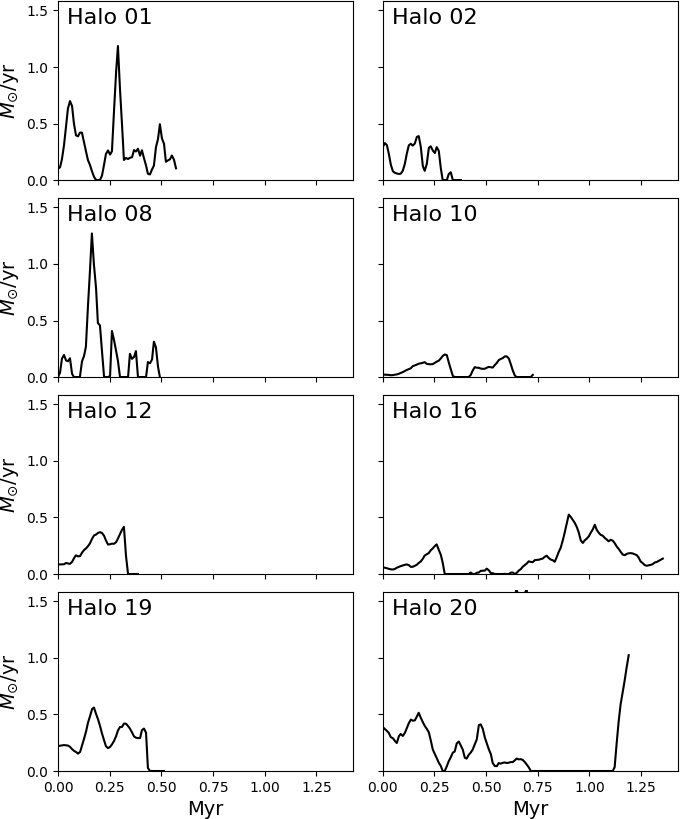}
    \caption{Accretion rates from \citet{pat20a} for the secondary stars. Note that times here are measured from the formation of the secondary disk, not the beginning of the simulation.}
    \label{fig:acc_hist}
\end{figure*}

The eight halos in \citet{pat20a} were chosen to span a range of assembly histories and spin parameters and had masses of $\sim$ 1 - 9 $\times$ 10$^7$ \Ms\ at collapse at redshifts $z =$ 13.9 - 20.4.  Resolution at the smallest scales on which fragmentation is known to occur in the disks had to be sacrificed in these simulations to follow their evolution for the times required for the stars to form, evolve and collapse \citep[e.g.,][]{bec15,bec18}.  Our accretion rates therefore exclude these fragments, but they are expected to fall onto the protostar prior to the main sequence \citep{ih14}.  We show accretion rates only for the largest companion disks that form in the halos in Figure~\ref{fig:acc_hist}, not smaller fragments that could undergo partial collapse but are short lived because they are taken up by the main disk or are tidally disrupted by it.

\subsection{\sc{kepler}}\label{sec:kepler}

We follow the evolution of the stars with {\sc kepler} \citep{kep1,kep2}.  {\sc kepler} typically partitions each star into a few thousand zones, over which it solves the energy and angular momentum equations: 

\begin{eqnarray}\label{cons_mom}
\frac{dv}{dt} & = & 4\pi r^{2} \frac{\partial P}{\partial m_{r}} + \frac{4\pi}{r}\frac{\partial Q}{\partial m_{r}} - \frac{G_{\rm{rel}}m_{r}}{r^{2}} \\
\frac{du}{dt} & = & -4\pi P \frac{\partial}{\partial m_{r}} (vr^{2}) + 4\pi Q \frac{\partial}{\partial m_{r}}\left(\frac{v}{r}\right) - \frac{\partial L}{\partial m_{r}} + \epsilon ~~~ 
\end{eqnarray}
where $r$ is the radius, $v$ is the velocity, $P$ is the pressure, $u$ is the internal energy per unit mass, $\epsilon$ is the local energy generation rate per unit mass, and $L$ is the rate of energy flow through a shell of radius $r$. We include a first-order post-Newtonian correction to the gravitational constant, $G_{\rm{rel}}$:
\begin{equation}
G_{\rm{rel}} = G \left(1 + \frac{P}{\rho c^{2}} + \frac{4\pi Pr^{3}}{m_{r}c^{2}}\right)\left(1 - \frac{2Gm_{r}}{rc^{2}}\right)^{-1}
\end{equation}
In Equation~\ref{cons_mom}, the factor $Q$ in the viscous term is
\begin{equation}
Q = \frac{4}{3}\,\eta_{\nu}r^4\frac{\partial}{\partial r}\left(\frac{v}{r}\right)\;,
\end{equation}
\noindent where $\eta _{\nu}$ is the dynamic viscosity as defined in \citet{kep1}:
\begin{equation}
    \eta _{\nu} = \eta _\mathrm{R} + \frac{3}{4}l_{1}\rho c_\mathrm{s} + \frac{3}{4}l_{2}^{2} \; \rm{max}(0, -\nabla \cdot \bf{v})
\vspace{0.1in}
\end{equation}
which includes both the real viscosity, $\eta _\mathrm{R}$, and the artificial viscosity, which can be modified arbitrarily to dampen acoustic oscillations during quiescent phases of the evolution of a star.  In this study, we take the standard values of $l_{1} = 0.1\Delta r$ and $l_{2} = 2\Delta r$, where $\Delta r$ is the width of a grid zone. At each time step, {\sc Kepler} initially attempts to make an arbitrarily large jump $\Delta t$ before iterating to find the maximum time step permitted by preset restrictions on the change in radius, temperature, density, luminosity, or velocity between zones, allowing us to model long-lived, quiescent evolutionary phases and to follow the emergence of shocks or the onset of collapse on short timescales.  The masses of the stars at collapse triggered by the the post-Newtonian instability \citep{chandra64} in {\sc Kepler} are consistent with analytic predictions \citep{Haemmerle20}.

To close these equations, nuclear-burning and energy generation are coupled to the hydrodynamics and solved with an adaptive network \citep{kep3}, convection is treated in a time-dependent manner with heat transport following the Ledoux criterion \citep{kep1}, and a Helmholtz-like equation of state is used that incorporates electron-positron pair production, relativistic and non-relativistic degenerate and non-degenerate electrons, and radiation \citep{ts00}. We neglect mass loss because wind and pulsational mass losses are expected to be negligible relative to the accretion rates in these stars \citep{vink01,bhw01,hos13}. For simplicity, we neglect rotation although it could affect the stellar structure \citep{hle18a}.

As in previous studies \citep{hos13, tyr17, hle18b, tyr21a}, we initialize all models as 10 \Ms, $n =$ 3 polytropes with central densities $\rho_{\mathrm{c}} =$ 10$^{-3}$ g cm$^{-3}$ and temperatures $T_{\mathrm{c}} =$ $1.2 \times 10^6\,$K, i.e., at the onset of deuterium burning.   We treat accretion onto the star and the associated advection of entropy as described in \cite{kep3} and \cite{tyr17}. The star is evolved until the onset of collapse, the end of the {\sc enzo} simulation, or if the evolution time corresponds to a moment in the {\sc enzo} simulation when the companion disk interacted strongly with or merged with the primary disk. 

\section{Evolution of Multiple Supermassive Stars in Primordial Halos}\label{sec:evolution}


\begin{figure*}
\begin{center}
\begin{tabular}{cc}\label{fig:kippenhahn} 
\stackinset{c}{}{t}{0cm}{\textbf{Halo 01}}{\epsfig{file=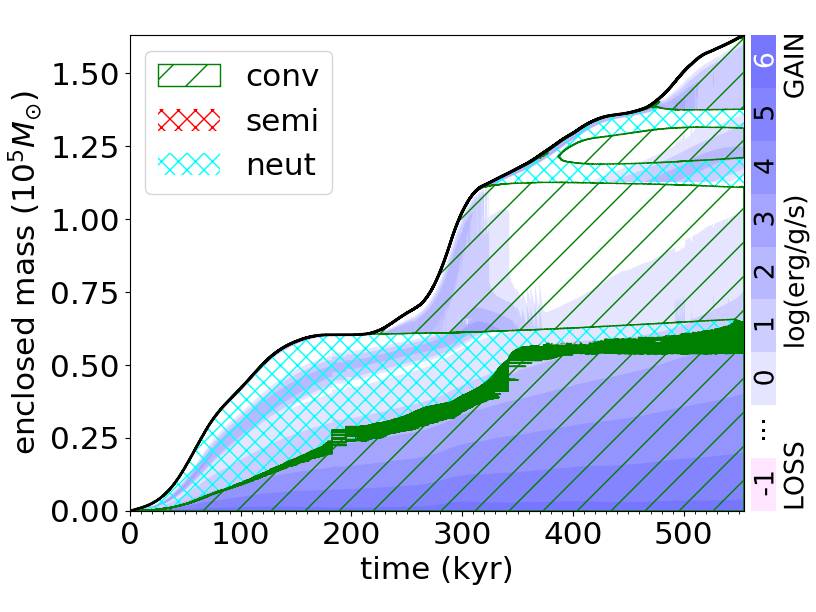,width=0.4\linewidth,clip=}} &
\stackinset{c}{}{t}{0cm}{\textbf{Halo 02}}{\epsfig{file=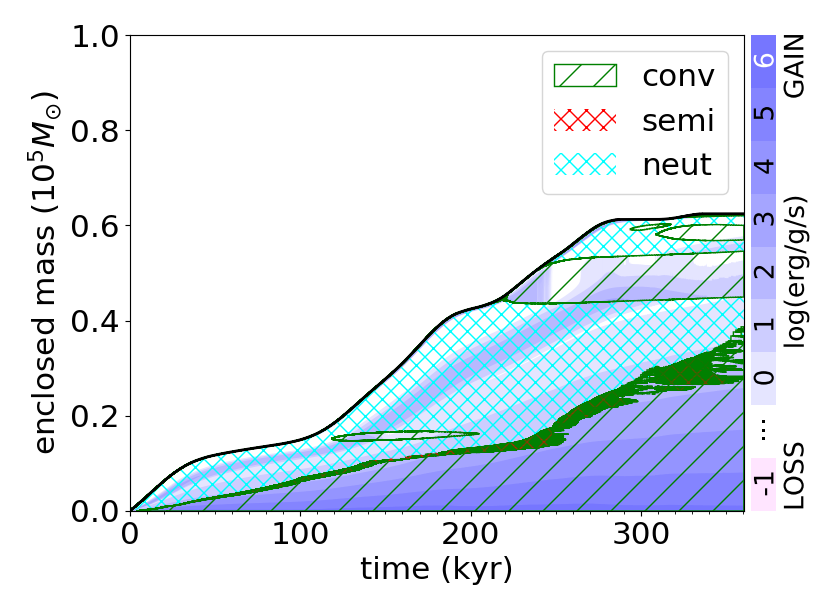,width=0.4\linewidth,clip=}}  \\
\stackinset{c}{}{t}{0cm}{\textbf{Halo 08}}{\epsfig{file=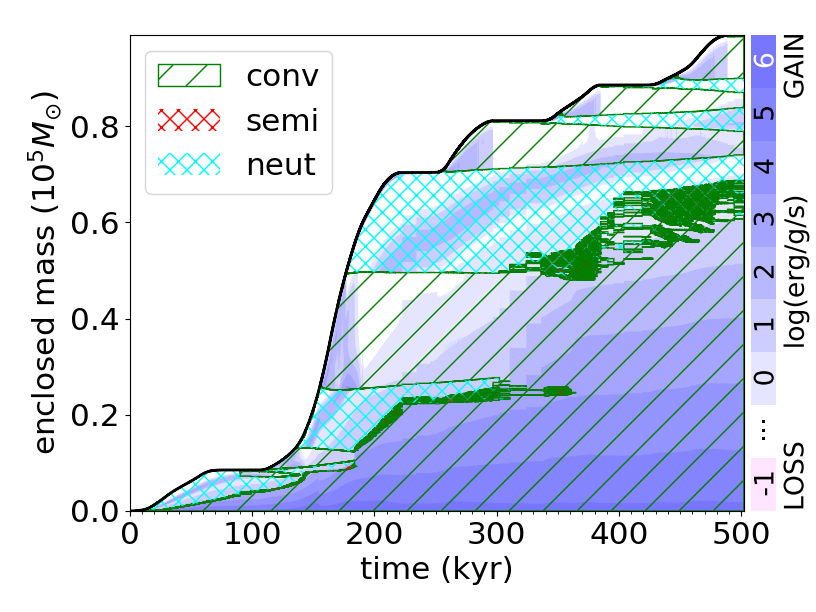,width=0.4\linewidth,clip=}}  &
\stackinset{c}{}{t}{0cm}{\textbf{Halo 10}}{\epsfig{file=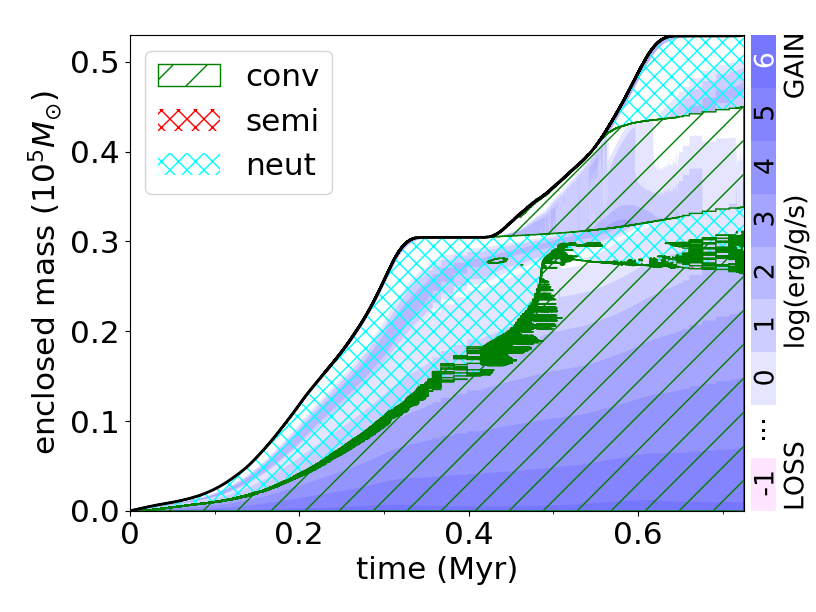,width=0.4\linewidth,clip=}}  \\
\stackinset{c}{}{t}{0cm}{\textbf{Halo 12}}{\epsfig{file=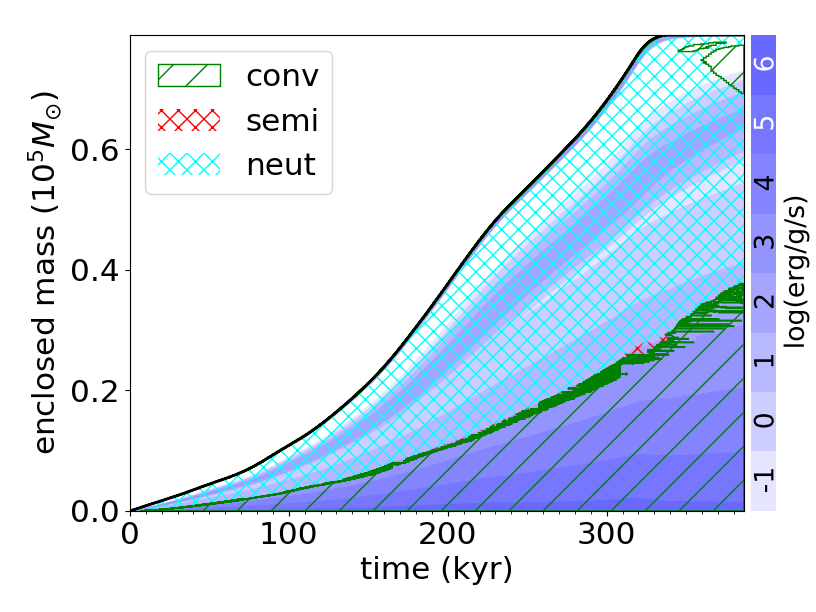,width=0.4\linewidth,clip=}} &
\stackinset{c}{}{t}{0cm}{\textbf{Halo 16}}{\epsfig{file=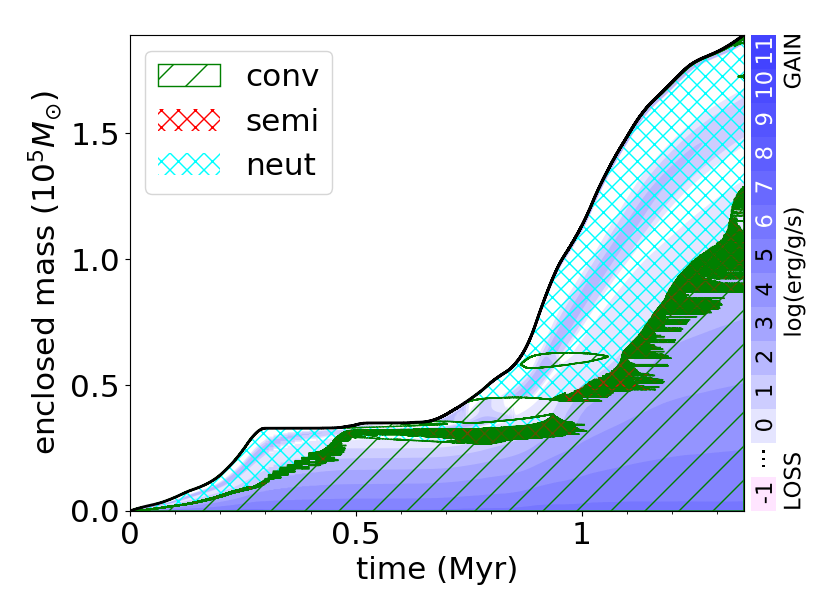,width=0.4\linewidth,clip=}}  \\
\stackinset{c}{}{t}{0cm}{\textbf{Halo 19}}{\epsfig{file=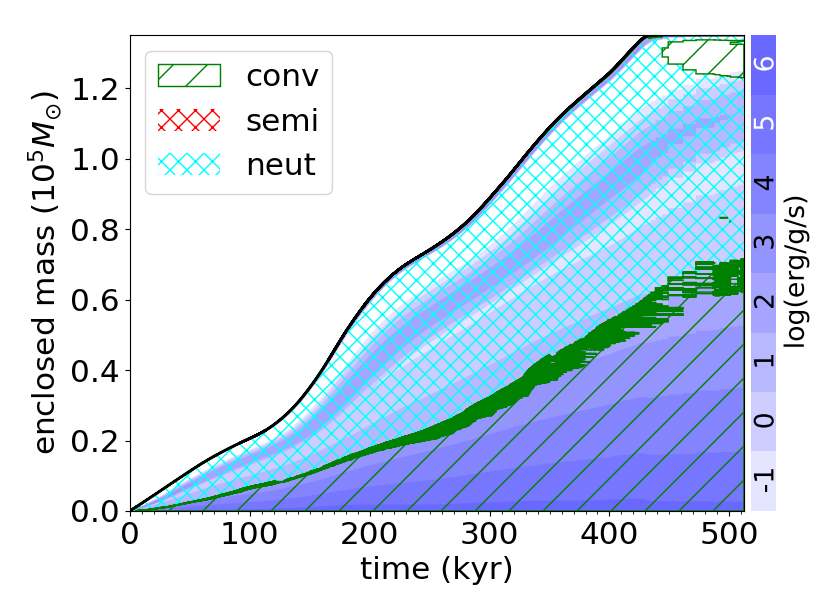,width=0.4\linewidth,clip=}}  &
\stackinset{c}{}{t}{0cm}{\textbf{Halo 20}}{\epsfig{file=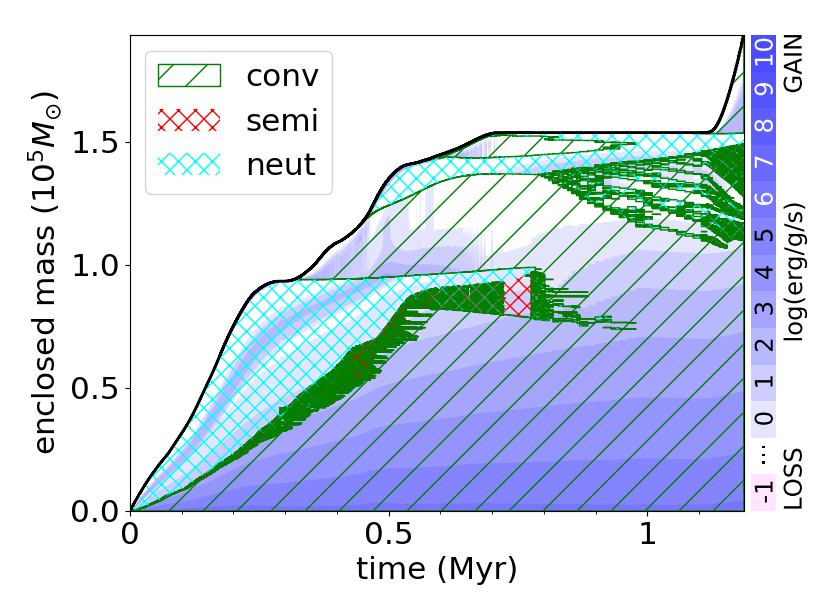,width=0.4\linewidth,clip=}}  \\
  \\
\end{tabular}
\caption{Kippenhahn diagrams of the internal structures of the stars in the satellite disks over time.  The blue colors indicate energy generation rates from nuclear-burning, green lines denote convective regions, red lines mark semi-convective regions, and light blue lines indicate radiative/convectively neutral regions.}
\label{fig:kip}
\end{center}
\end{figure*}

\begin{table*}[b]
\centering
\begin{tabular}{c|c|c|c|c|c|c|c|c|c}
\hline\hline
 Halo & $M_{1}$ & $t_{sim}$ & Evol Stage (1) & $X_{c,1}$ & $M_{2}$ & $t_{2}$ & Evol Stage (2) &  $X_{c,2}$ & Close Pair? \\
         & k\Ms\ & Myr & &  &  k\Ms\ & Myr & &  & \\
        \hline
        01 & 134 & 0.947 & MS & 0.38 & 165 & 0.571 & MS & 0.40 &  yes \\
        02 & 143 & 1.46 & BH & 0.34 &  62 & 0.378 & MS & 0.58 & yes \\
        08 & $\geq$186 & 1.014 & BH & --- &  99 & 0.503 & MS & 0.56 &  yes\\
        10 & $\geq$132 & 2.006 & BH & --- &  53 & 0.639 & MS & 0.55 &  no \\
        12 & $\gtrsim$178 & 1.395 & BH & --- &  79 & 0.380 & MS & 0.57 &  yes \\
        16 & $\geq$109 & 3.054 & BH & --- &  186 & 1.27 & MS (near collapse?) & 0.39 &  no\\
        19 & 46 & 1.439 & MS & 0.28 &  133 & 0.426 & MS & 0.54 &  yes\\
        20 & $\geq$178 & 1.770 & BH & --- &  154 & 1.103 & MS (near collapse?) & 0.34 &  no \\
        \hline\hline
\end{tabular}
\caption{Final masses, ages, evolutionary stages (ES), and central hydrogen fractions for the primary (1) and secondary (2) SMSs in each halo, followed by whether or not they form a close pair by the end of the simulation.  Note that the SMS in the primary disk in \textbf{Halo 02} collapses to a black hole during the merger with the secondary disk. \label{tab:my_label}}
\end{table*}

We find that all the secondary disks in the halos in \citet{pat20a} form long-lived nuclear-burning stars. Like the stars in the primary disks \citep{tyr21a}, we find a striking variety of internal structures for these secondary SMSs as shown in Figure~\ref{fig:kip}, in marked contrast to stars that evolve under constant accretion rates \citep[e.g.,][]{hos13, tyr17, hle18b}.  The stars all initially exhibit a deep radiative envelope corresponding to the surge in accretion associated with the formation of their natal disks, which proceeds on much shorter times than the star's thermal timescale and leads to the buildup of a steep entropy gradient \citep[e.g.,][]{begel10,hos13,tyr17,hle18b}. In several cases, such as {\bf Halos 01} and {\bf 16}, however, almost all of mass of the star eventually lies within its convective core because there is a long ($\gtrsim$ 100 kyr) quiescent phase ($\gg$ Kelvin-Helmholtz time-scale) in accretion in which the star can thermally relax and its structure can approach that of an $n=3$ polytrope \citep{chandra64, tyr20a}.  In other models, we see the formation of both transient and long-lived convective cells in the otherwise deep, high-entropy radiative envelopes of some rapidly-accreting stars, similar to those in some constant-accretion rate models \citep[e.g.,][]{tyr17, hle18b}.

In \textbf{Halo 01}, a secondary disk forms at 400 kyr by breaking off from one of the spiral arms of the primary disk. Throughout its evolution, this companion disk remains in an elliptical orbit around the first with typical separations of 0.5 - 1 pc.  The sharp spike in accretion $\sim$300~kyr after formation is due to a merger with a clump in the halo. Accretion continues until the second disk merges with the first 650 kyr after formation.  At this time, the SMS in the second disk is a somewhat evolved main sequence star with a central hydrogen fraction of $\sim$0.4. The SMS in the first disk is only slightly more evolved, with a central hydrogen fraction of 0.38.  Although the subsequent evolution of these stars is beyond the scope of this work, \citet{tyr21a} assumed that the collision of the disks would trigger a surge in accretion onto the first star but the presence of the second SMS may lead to other outcomes, as discussed in $\S$ \ref{MSMS}.

\textbf{Halo 02} is an example of a particularly turbulent, chaotic system in which three  clumps form and interact with the primary and companion disks \citep{pat20a}.  Interactions between these clumps and the second disk lead to two large bursts of accretion at 130 kyr and 250 kyr that create a SMS that grows to $\mathord\sim60$ k\Ms\ by 275 kyr and has a deep radiative envelope.  At this point, there is a brief quiescent phase in accretion due to a close encounter between the two disks.The two disks then merge 378 kyr after the formation of the second disk. The first and second SMSs are 62 k\Ms\ and 143 k\Ms, respectively, and have central hydrogen fractions of 0.58 and 0.34.  Notably, the collision of the disks produces a large spike in accretion onto the first SMS that quickly brings it up to the post-Newtonian instability and causes it to collapse \citep{tyr21a}. 

Only one satellite disk forms in \textbf{Halo 08}, which lasts from 685 kyr to about 1 Myr after the formation of the first disk. Accretion onto the star is highly variable because of the eccentric orbit of the satellite disk, and the fluctuations correlate with the smallest and largest distances between the two disks, which are  $\sim$ 0.3 pc and 2.0 pc, respectively.  This accretion history produces the distinctive step-like structure in the Kippenhahn diagram of the star in Figure~\ref{fig:kip}.  About 1 Myr after the formation of the primary disk, the star in the second disk has an age of 503 kyr, a mass of 99 k\Ms, and a central hydrogen fraction of 0.56. The SMS in the primary disk, however, has by this time collapsed to a black hole, having encountered the post-Newtonian instability at an age of 950 kyr when it reached a mass of 186 k\Ms\ and a central hydrogen fraction of 0.38.  

The first disk in \textbf{Halo 10} is stable and does not begin to fragment for 1.14 Myr.  The second disk forms $\sim$ 1.2 pc from the center of the first in an initially highly elliptical orbit.  Two major episodes of accretion drive the growth of the second SMS that in each case build up a deep, high-entropy envelope.  Between these episodes there is a long (nearly 100 kyr) quiescent phase during which the star thermally relaxes without becoming entirely convective.  This star reaches a  mass of 53\ k\Ms\ and a central hydrogen fraction of 0.55 by the end of the simulation.  At this time the secondary disk is still in an elliptical orbit around the first with separations that vary from 0.5 -- 1 pc. The SMS in the first disk, however, has collapsed because of the post-Newtonian Chandrasekhar instability late on the main sequence at an age of 1.95 Myr, a final mass of 132 k\Ms, and a central hydrogen fraction of 0.06.  

The primary disk in \textbf{Halo 12} is highly turbulent and frequently fragments, but most of the clumps soon migrate to the center of the disk. The longest-lived of the clumps forms at $\sim$ 1 Myr and produces its own disk that survives for 380 kyr. Rapid accretion in this disk creates a SMS with a deep radiative envelope that grows to a mass 79 k\Ms\ with a central hydrogen fraction of 0.57 by the time the two disks merge. About 200 kyr before the destruction of the second disk, the SMS in the primary disk collapsed via the Chandrasekhar instability while still on the main sequence at a final mass of 178 k\Ms\ and central hydrogen fraction of 0.32. 

Like \textbf{Halo 10}, \textbf{Halo 16} has a relatively stable disk that does not fragment for $\sim$ 1.5 Myr.  A key difference, however, is that the SMS in the primary disk collapses to a BH at about the time the second disk forms because it encounters the post-Newtonian instability at the very end of the main sequence at a mass of 109 k\Ms, when its core hydrogen is exhausted.  It is unclear from the Enzo simulation how X-rays from this black hole would affect the evolution of the companion disk but we discuss some possible outcomes in $\S$\ref{outcomes}.  As the primary disk continues to fragment, 3-body interactions fling the companion disk into a highly elliptical orbit $\sim$ 300 kyr after its formation.  The initial burst of accretion is followed by a long quiescent phase ($\sim$ 0.5 Myr) during which the second SMS, which is now $\sim$ 40 k\Ms, becomes almost completely thermally relaxed and almost fully convective \citep{tyr20a}. The rapid accretion beginning at $\sim$ 750 kyr builds up a massive radiative envelope on top of this convective core and the mass of the star grows to 186 k\Ms\ by the end of the simulation.  At this point the SMS is still on the main sequence in a disk on a long orbit around the first disk, but it appears to be on the verge of collapse.  With a mass nearing the upper limit for SMSs with similar accretion rates \citep[e.g.,][]{tyr17, hle18b, tyr21a}, the star is unlikely to survive for much longer so the system may soon produce a DCBH binary. 

As the primary disk in \textbf{Halo 19} begins to fragment, two clumps merge and form a stable companion disk after $\sim$ 800 kyr.  The initial burst of accretion due to the formation of this disk is particularly strong and builds up a star with a massive convective envelope in $\sim$ 500 kyr.  Accretion in the disk is so rapid that the evolution of the second SMS overtakes that of the first.  By the time the disks merge, the first SMS has reached a mass of 46 k\Ms\ and a central hydrogen fraction of 0.28 but the second is at a mass of 133 k\Ms\ and a central hydrogen fraction of 0.54. Both stars are still on the main sequence.

The primary disk in \textbf{Halo 20} begins to fragment soon after formation and quickly forms a single, massive companion disk. Accretion rates in the disk are fairly high, and are rejuvenated at one point when a clump collides with the disk. These high rates quickly build up a SMS with a deep, high-entropy envelope. The companion disk later exchanges mass with the primary disk over a number of orbits that mostly halts accretion onto the second star and allows it to thermally relax. In the meantime, the SMS in the primary disk collapses at 1.48 Myr to a BH via the post-Newtonian/Chandrasekhar instability late in the main sequence at a core hydrogen fraction of 0.14 and a mass of 189 k\Ms.  At 1.77 Myr, the end of the simulation, the companion disk is in a relatively long elliptical orbit ($\sim$ 1.8~pc separation) that is growing in radius.  At this point, the companion is still on the main sequence but, as in \textbf{Halo 16}, appears to be about to collapse, having reached a mass of 154 k\Ms.

\section{Discussion}\label{sec:interactions}

In most of the cases above, the accretion history and evolution of the second SMS ends with the merger of its host disk with the primary disk. Although the subsequent evolution of the stars or DCBHs in the disks cannot be determined here, a number of outcomes can be inferred for them for follow-up with future simulations, as we outline below.

\subsection{MS-MS Pairs}
\label{MSMS}
The primary and secondary disks in \textbf{Halo 01}, \textbf{Halo 19}, and nominally \textbf{Halo 02} merge while their central stars are still on the main sequence (MS).  With a maximum physical resolution of 0.014 pc ($\sim$ 3 kAU), the Enzo simulations cannot determine if they later interact.  Drag forces in the inner disk could bring about a swift merger between the two SMSs or the stars could carve out a gap in the circumbinary disk.  Just before the merger of the two disks in \textbf{Halo 02} one of the SMSs appears to be on the verge of collapse via the post-Newtonian instability. Examples such as this make it clear that a more careful treatment of stellar structure during the resulting enormous surge in accretion is essential \citep{MH17} and could be informed by smoothed-particle hydrodynamics calculations \citep[e.g.,][]{Glebbeek13}. Such mergers may also produce unique observational transients that signal the formation of SMSs in the early Universe. If instead these supermassive stellar pairs evolve into relatively long-lived binaries, they could later become BH -- MS or BH -- BH pairs. 

\subsection{MS-BH Pairs}
\label{outcomes}
In \textbf{Halos 08 and 12} the primary and companion disks merge after the star in the primary disk has collapsed to a DCBH while the SMS in the other disk is still on the main sequence. Given that the masses of both objects are comparable in both halos, later interactions between the two are unlikely to produce tidal disruption events. The tidal disruption radius
\begin{equation}
    R_{\rm t} = R_{*}\left(\frac{M_{\rm BH}}{M_{*}}\right)^{\frac{1}{3}}
\end{equation}
in such cases is of the order of the stellar radius and would therefore only happen during a collision. However, fragmentation on scales smaller than those resolved here may lead to the formation of much less massive stars that could then be torn apart by the DCBH and produce a powerful radio or NIR transient \citep{ki16,Regos2021}.

If these objects form a long-lived binary, they will continue to evolve subject to additional torques from the circumbinary disk. Depending on the orbital evolution, the companion SMS may overflow its Roche lobe while still on the main sequence or as it evolves, in which case its deep radiative envelope could sustain  long-lived mass exchange and, perhaps, the formation of a ``supermassive'' X-ray binary \citep[e.g.,][]{Soberman1997}. Eventually, the collapse of the second SMS due to exhaustion of core nuclear fuel or the Chandrasekhar instability will produce a DCBH binary as discussed below.

\subsection{BH-BH Pairs}

\begin{figure}
    \centering
    \includegraphics[width=\columnwidth]{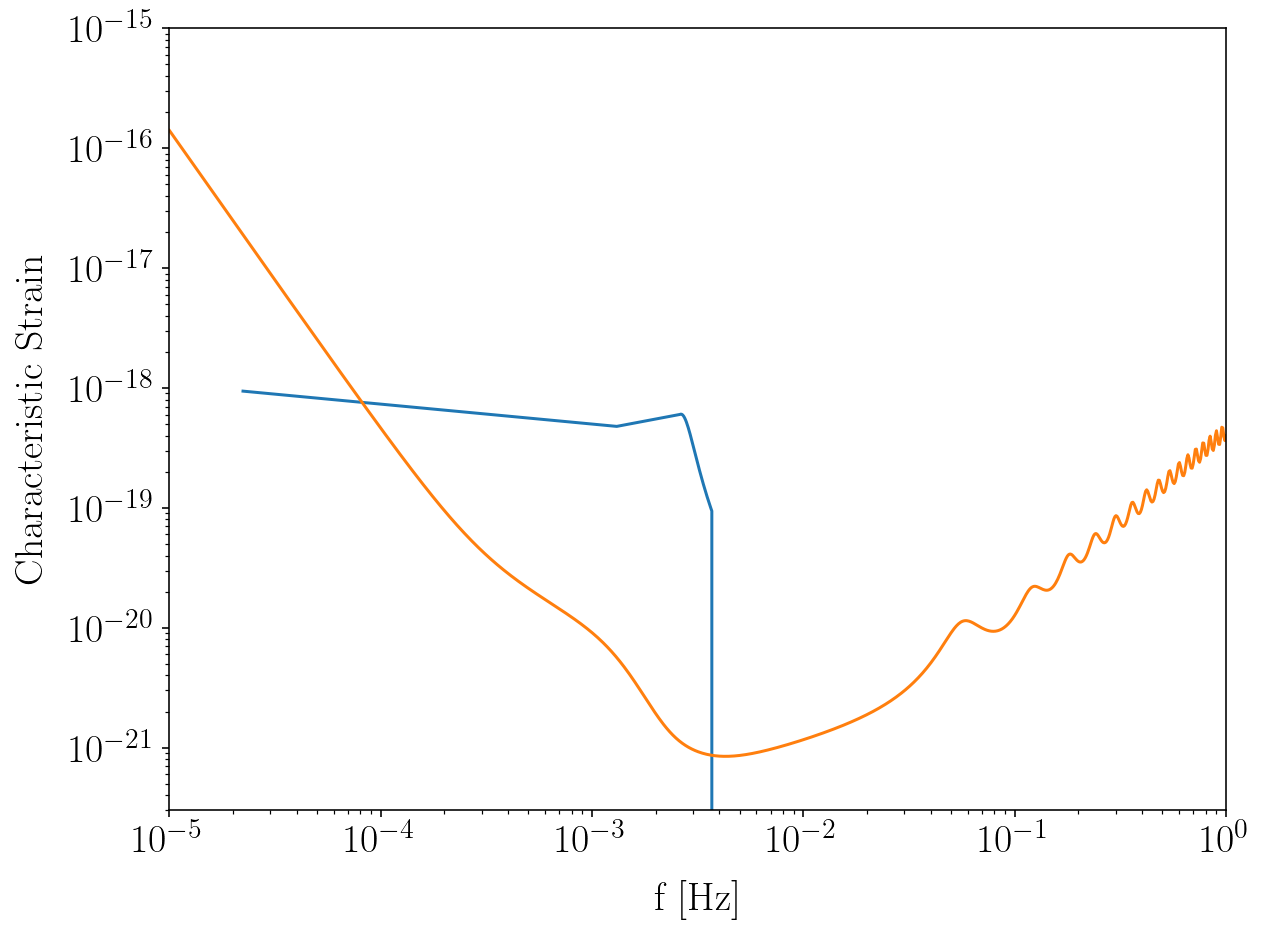}
    \caption{Sky-averaged LISA sensitivity curve (orange) and the observable characteristic strain (blue) from the merger of the DCBHs in \textbf{Halo 20}.}
    \label{fig:GWstrain}
\end{figure}

Some of our models may produce long-period DCBH binaries.  At the end of the simulation in \textbf{Halo 10} the second SMS was still on the main sequence but its host disk was in a highly elliptical orbit with the first disk, whose star had collapsed to a DCBH.  This  long-period orbit suggests that the SMS will collapse to a DCBH before interacting closely with the other BH.  Although the secondary SMSs in \textbf{Halos 16} and \textbf{20} are also still on the main sequence at the end of the simulation they are both on the verge of collapse via the post-Newtonian instability unless accretion stops. Both of these halos could thus produce long-period ($\sim 1$ pc separation) binary DCBHs. The timescale for a merger between the two BHs would depend on their dynamics in the halo and drag forces due to gas flows therein, and the strength of their gravitational wave (GW) signal depends on their masses when they finally do merge.  Here, we estimate the GW emission from the merger of the DCBH binary in \textbf{Halo 20} \citep{LISAcurves}\footnote{LISA Sensitivity Calculator available for download from:\\ \url{github.com/eXtremeGravityInstitute/LISA_
Sensitivity.git}}, considering a sky and polarization averaged transfer function and assuming that the orbit has circularized by prior GW emission. The expected signal is shown in Figure~\ref{fig:GWstrain} for the redshift at which the halo collapses, $z = 17.7$.  This merger will be detectable by {\it LISA}, as would mergers of the other BH -- BH pairs in our halos at their respective redshifts.

\section{Conclusions}

A growing number of recent numerical simulations indicate that the formation of multiple very massive stars \citep[e.g.,][]{wise19, Regan20} or SMSs \citep[e.g.,][]{latif20a,pat20a} was common in primordial, atomic-cooling halos, and that these could be the origin of the first quasars in the Universe. Here, we have shown that secondary disks in these halos form SMSs whose accretion histories can be coupled to those of the stars in the primary disks via tidal interactions.  Most satellite disks merge with the primary disk in just 1 -- 2 Myr, which could yield supermassive stellar mergers, supermassive X-ray binaries, massive black hole seed mergers, and SMS binaries.  Such interactions could profoundly affect the evolution of the stars, the SMBH seeds they produce, and their observational signatures.  Simulations that follow the dynamics of these binaries and mass exchange between them with stellar evolution calculations will be the aim of future studies.

\begin{acknowledgments}

T.E.W.\ acknowledges support from the National Research Council Canada's Plaskett Fellowship. S.P.\ was supported by STFC grant ST/N504245/1 and D.J.W.\ was supported by the Ida Pfeiffer Professorship at the Institute of Astrophysics at the University of Vienna.  A.H.\ was supported by the Australian Research Council (ARC) Centre of Excellence (CoE) for Gravitational Wave Discovery (OzGrav) through project number CE170100004, by the ARC CoE for All Sky Astrophysics in 3 Dimensions (ASTRO 3D) through project number CE170100013, and by the National Science Foundation under Grant No. PHY-1430152 (JINA Center for the Evolution of the Elements, JINA-CEE).

\end{acknowledgments}

\bibliography{sample631}{}
\bibliographystyle{aasjournal}

\end{document}